\documentclass[prl,twocolumn,superscriptaddress,usenames,dvipsnames]{revtex4-2} 

\usepackage{amsmath,bm}
\usepackage{csquotes}
\usepackage{hyperref}
\usepackage{graphicx}

\usepackage{color}
\graphicspath{{img/}}
\newcommand{\vecg}[1]{\boldsymbol{#1}}

\usepackage[normalem]{ulem}

\begin{document}

\title{Coexistence of uniform and oscillatory states resulting from nonreciprocity and conservation laws}

\author{Daniel Greve}  
\email{daniel.greve@uni-muenster.de}
\affiliation{Institut für Theoretische Physik, Universität Münster, Wilhelm-Klemm-Str. 9, 48149 Münster, Germany}

\author{Giorgio Lovato}  
\email{stu242274@mail.uni-kiel.de}
\affiliation{Institut für Theoretische Physik und Astrophysik, Christian-Albrechts-Universität zu Kiel, Leibnizstr. 5, 24098 Kiel,Germany}

\author{Tobias Frohoff-Hülsmann}
\email{t\_froh01@uni-muenster.de}  
\affiliation{Institut für Theoretische Physik, Universität Münster, Wilhelm-Klemm-Str. 9, 48149 Münster, Germany}

\author{Uwe Thiele}  
\email{u.thiele@uni-muenster.de}
\homepage{www.uwethiele.de}
\affiliation{Institut für Theoretische Physik, Universität Münster, Wilhelm-Klemm-Str. 9, 48149 Münster, Germany}
\affiliation{Center for Nonlinear Science (CeNoS), Universität Münster, Corrensstr. 2, 48149 Münster, Germany}
\affiliation{Center for Multiscale Theory and Computation (CMTC), Universität Münster, Corrensstr. 40, 48149 Münster,Germany}

\begin{abstract}
  Employing a two-species Cahn-Hilliard model with nonreciprocal interactions we show that the interplay of nonreciprocity and conservation laws results in the robust coexistence of uniform stationary and oscillatory phases as well as of uniform and crystalline phases. For nonequilibrium models with a spurious gradient dynamics structure such coexistencies between two or more nonequilibrium phases and resulting phase diagrams can nevertheless be predicted by a Maxwell double-tangent construction. This includes phases with sustained regular or irregular out-of-equilibrium dynamics as further corroborated by bifurcation studies and time simulations. 
\end{abstract}

\maketitle

Thermodynamic out-of-equilibrium processes like phase separation are often described by gradient dynamics models, i.e., with continuum theories for the overdamped time evolution of density-like order parameter fields. The existence of an underlying thermodynamic potential results in a monotonic relaxation toward an equilibrium state. Mass conservation may imply that different phases coexist in extensive parameter ranges, e.g., for a decomposing mixture as described by the classical Cahn-Hilliard (CH) model~\cite{CaHi1959jcp}. In the thermodynamic limit the coexisting phases are predicted by a common-tangent Maxwell construction~\cite{Maxw1875nat}. 

Recently, active mixtures that remain permanently out of equilibrium, e.g., due to an underlying chemo-mechanical coupling, have gained much attention. Descriptions by continuum theories arise from coarse-graining the dynamics of active particles \cite{DOCT2023nc,CaTa2015arcmp,VrBW2023jpcm}, as phenomenological models \cite{SaAG2020prx,YoBM2020pnasusa, FrWT2021pre}, and as amplitude equations \cite{FrTh2023prl,BeRZ2018pre,RaBZ2019epje}. Examples like active model B (AMB) \cite{WTSA2014nc,TjNC2018prx,BiWi2020prr} and the nonreciprocal Cahn-Hilliard (NRCH) model \cite{YoBM2020pnasusa,SaAG2020prx,FrWT2021pre,SuKL2023prl} correspond to nonvariational generalizations of CH models. Thereby,  AMB arises for motility-induced phase separation \cite{CaTa2015arcmp} while NRCH models describe active ternary or higher order mixtures. Further, the latter capture universal large-scale oscillatory dynamics for systems with two conservation laws~\cite{FrTh2023prl}.
Remarkably, for the single-species AMB case a Maxwell-like construction predicts coexisting densities even though the system is nonvariational~\cite{WTSA2014nc,SSCK2018pre}. 

\begin{figure}[tb]
\centering
\includegraphics[width=1.05\hsize]{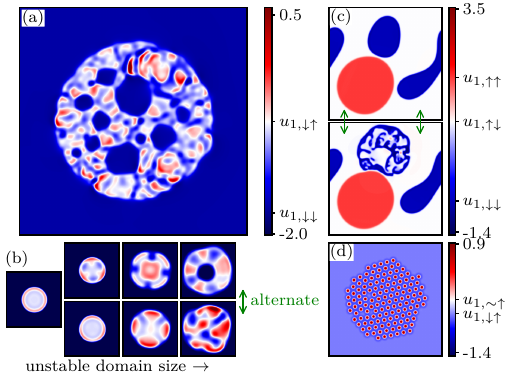}
\caption{Predicted phase coexistences in active phase decomposition [NRCH model \eqref{eq:nrch}]. Panel (a) displays a snapshot of the density $u_1$ for the coexistence of a central cluster of an oscillatory phase with irregular wave dynamics about the high-density state $u_{1,\downarrow\uparrow}$ and a uniform low-density state $u_{1,\downarrow\downarrow}$ (blue). Panel~(b) illustrates the transition from (left) a steady state to (middle) regular, and (right) irregular wave dynamics with increasing size of the central domain for otherwise identical parameters as in (a). Panel~(c) shows three-phase coexistence (red/blue/white) using two snapshots that illustrate (top) a \enquote{calm} period and (bottom) one of the occasional burst of wave activity in the $\downarrow\downarrow$-phase (blue). Panel~(d) shows a central cluster of a stationary crystalline phase that coexists with a uniform phase ($u_{1,\downarrow\uparrow}$). Coexisting (mean) densities in in (a,b), (c) and (d) are marked by square symbols in Figs.~\ref{fig:phase_diagrams}(b) to~(d), respectively. The parameters are (a-d) $a=-1.5$, $\rho=1$, (a,b) $\zeta=3, \kappa=1$, (c) $\zeta=0.69,\kappa=1$ and (d) $\zeta=8,\kappa=20$.
 The domain size in (a), (c) and (d) is $80\pi \times 80\pi$, and in (b) $30\pi \times 30\pi$ (only $15\pi \times 15\pi$ is shown). 
For numeric details and accompanying movies see SM.}
\label{fig:2dcoexistence}
\end{figure} 

Here, we explain why for an entire third class of multi-species models between classic passive (thermodynamic) ones and fully active ones, generalized Maxwell-type constructions predict nonequilibrium phase diagrams in the thermodynamic limit. We show that this is a further consequence of their spurious gradient dynamics form as introduced in Ref.~\cite{FHKG2023pre} to explain the unexpected existence of steady asymmetric states in several common nonvariantional models. This third class contains, e.g., selected active phase-field-crystal (aPFC) \cite{MeLo2013prl,OpGT2018pre} and NRCH models. In consequence, for the entire model class one can even capture the behavior of nonequilibrium phases that only exist due to activity. Figs.~\ref{fig:2dcoexistence}(a-c) and Fig.~\ref{fig:2dcoexistence}(d) illustrate the predicted coexistences of uniform and oscillatory states and of uniform and crystalline states~\footnote{Due to their analogy to PFC structures, we refer to crystal-like periodic microphase-separated structures as \enquote{crystalline} and to corresponding localized states as \enquote{crystallites}}. Fig.~\ref{fig:2dcoexistence}(a) gives a large-scale impression of two-phase coexistence where an oscillatory phase shows irregular waves while Fig.~\ref{fig:2dcoexistence}(b) illustrates the sequence of egular and irregular wave patterns at small sizes of the oscillatory patch [cf.\ Movies~1 and 2 in the Supplementary Material (SM)\footnote{See Supplementary Material at \url{http://link.aps.org/
supplemental/10.1103/PhysRevLett.134.01830} which includes Refs.~\cite{HoAT2021jpcm, Turi1952ptrslsbs, UeWR2014nmma, Ueck2019ccp,GLFT2024zenodo} for additional information and details on the numerical methods used.}].
Fig.~\ref{fig:2dcoexistence}(c) showcases that even three-phase coexistence of two uniform and one oscillatory phase can be predicted by the Maxwell construction.

\begin{figure*}[htb]
\includegraphics[width=\hsize]{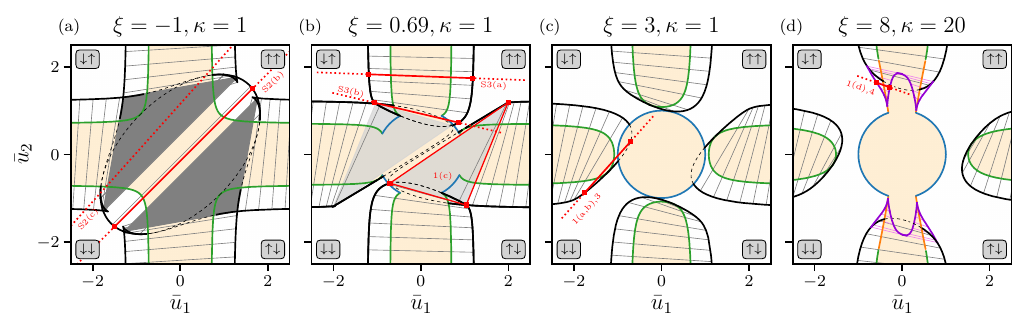}
\caption{Phase diagrams in the ($\bar{u}_1,\bar{u}_2$)-plane for the NRCH model for increasing nonreciprocity $\xi=\alpha^2-\rho^2$ (left to right) at (a)-(c) unity ($\kappa=1$) and (d) nonunity ($\kappa=20$) rigidity ratio. For coexisting uniform phases, solid thick black and thin gray lines represent (stable and metastable) binodals and tie lines, respectively. Purple lines in (d) represent binodals for coexisting uniform and crystalline phases (tie lines are pink). Dashed black lines indicate unstable binodals, i.e., the coexisting state is unstable.  The dark [light] gray area in (a) [(b)] marks stable [unstable] three-phase coexistence. The white areas in the four corners are single-phase regions, e.g., top left of the $\downarrow\uparrow$-phase (low $u_1$, high $u_2$). Green, orange and blue lines indicate spinodals for Cahn-Hilliard, conserved-Turing and (oscillatory) conserved-Hopf instabilities, respectively. Remaining parameters are $\rho=1$ and $a=-1.5$. For red straight lines see main text.}

\label{fig:phase_diagrams} 
\end{figure*}

Here, the Maxwell construction for spurious gradient dynamics models and the resulting phase behavior is illustrated employing the linearly coupled two-species NRCH model \cite{FrWT2021pre}
\begin{equation}\label{eq:nrch}
\partial_tu_i=\vec\nabla \cdot \left[Q_{i} \vec \nabla \left(\frac{\delta\mathcal{F}}{\delta u_i} + \mu_{i}^\mathrm{nr}\right)\right].
\end{equation}
For $i=1,2$ the two conservation laws describe a mixture of nonreciprocally interacting species of densities $u_1(\vec{x},t)$ and $u_2(\vec{x},t)$ where the nonequilibrium chemical potentials are $\mu_{1}^\mathrm{nr}=-\alpha u_2$ and $\mu_{2}^\mathrm{nr}=+\alpha u_1$.
The underlying energy is $\mathcal{F}=\mathcal{F}_1[u_1]+\mathcal{F}_2[u_2]+\mathcal{F}_{12}[u_1,u_2]$ with $\mathcal{F}_{12}=\int_\Omega f_{12}\mathrm{d}^n x$ where $f_{12}=-\rho u_1 u_2$, and $\mathcal{F}_{i}=\int_\Omega[(\kappa_i/2)|\vec\nabla u_i|^2+f_i]\mathrm{d}^n x$ where $f_i=(a_i/2)u_i^2+(b_i/4)u_i^4$. The two $Q_{i}$ represent diffusional mobilities~\footnote{Cross-diffusion and dependence on $u_1$ and $u_2$ are neglected, as no change in our main argument would result.}. By scaling space, time and fields we set $b_1, b_2, \kappa_1$ and $Q_{1}$ to one. Then, $\kappa=\kappa_2/\kappa_1$ represents the rigidity ratio. For simplicity, we set $a_1=a_2=a$ and $Q_{2}=1$, and $n$ is the number of spatial dimensions. Reciprocal and nonreciprocal interactions are parametrized through respective symmetric ($\rho$) and antisymmetric ($\alpha$) coupling strengths.

The nonreciprocity parameter $\xi=\alpha^2-\rho^2$ distinguishes cases of dominant reciprocal ($\xi<0$) and dominant nonreciprocal ($\xi>0$) coupling.  For purely reciprocal coupling ($\alpha=0$), Eqs.~\eqref{eq:nrch} represent a proper gradient dynamics, see \cite{CaHi1958jcp,Cahn1965jcp,HoHa1977rmp,Bray1994ap,FrWT2021pre}.

The nonreciprocal interaction breaks Newton's third law, similar to a predator-prey attraction-repulsion interaction. As a result, for dominant nonreciprocity ($\xi>0$) the NRCH model features conserved-Hopf and -Turing instabilities \cite{FrTh2023prl,FrTP2023ptrsa}, suppression of coarsening \cite{FrWT2021pre}, localized patterns \cite{FrTh2021ijam}, sustained traveling/standing waves \cite{YoBM2020pnasusa,FrWT2021pre,SuKL2023preb,BrMa2024prx}, and more complex spatiotemporal patterns \cite{SaAG2020prx,FrWT2021pre} that are all not possible in passive CH models. The conserved-Hopf instability results in large-scale oscillatory behaviour as in the central clusters in Figs.~\ref{fig:2dcoexistence}(a,c). The conserved-Turing instability only occurs for unequal rigidities ($\kappa\neq 1$)  \cite{FrTP2023ptrsa} and gives rise to small-scale stationary patterns, see Fig.~\ref{fig:2dcoexistence}(d). Dispersion relations and spinodals are determined in section~S2 of the SM.

At first sight, one might assume that only in the passive case (${\alpha=0}$) coexisting states (binodals) can be obtained via a common-tangent Maxwell construction. However, for Eqs.~\eqref{eq:nrch} this holds even for dominant nonreciprocal coupling ($\xi>0$). Specifically, introducing the spurious energy
	 $\tilde{\mathcal{F}}=\frac{\rho}{\rho+\alpha}\mathcal F_1+\frac{\rho}{\rho-\alpha}\mathcal F_2+\mathcal{F}_\mathrm{12}$
and mobilities
	$\tilde M_{1}=\frac{\rho+\alpha}{\rho}$ and $\tilde M_{2}=\frac{\rho-\alpha}{\rho}$, Eqs.~\eqref{eq:nrch} can be written as a \textit{spurious gradient dynamics}. It is spurious as for $\xi>0$ the energy $\tilde{\mathcal{F}}$ is not bounded from below and the mobilities $\tilde M_{i}$ are not both positive, i.e., basic thermodynamic principles are not fulfilled any more. Nevertheless, as we show next, a resulting \textit{spurious Maxwell construction} allows us to obtain coexisting states, i.e., binodals and tie lines, from a double-tangent construction on $\tilde{\mathcal{F}}$. These have to be carefully scrutinized as in the nonreciprocal case, the binodals only represent part of the picture. Taken in combination with linear stability results, the spurious Maxwell construction allows us to obtain nonequilibrium phase diagrams that, e.g., predict the coexistence in Fig.~\ref{fig:2dcoexistence}.

Note that the resulting phase diagram features the spinodals and binodals for all uniform and crystalline phases. For two coexisting uniform phases A and B the double-tangent construction corresponds to the conditions $\tilde{\mu}_1(\vecg u^{(A)})=\tilde{\mu}_1(\vecg u^{(B)})$, $\tilde{\mu}_2(\vecg u^{(A)})=\tilde{\mu}_2(\vecg u^{(B)})$, and $\tilde p(\vecg u^{(A)})=\tilde p(\vecg u^{(B)})$
  where $\vecg u^{(A)}$ and $\vecg u^{(B)}$ are the respective densities, the spurious chemical potentials are $\tilde{\mu}_i=\partial \tilde f/\partial u_i$ with $\tilde f=\frac{\rho}{\rho+\alpha}f_1+\frac{\rho}{\rho-\alpha}f_2+f_{12}$, and the spurious pressure is $\tilde p=\tilde{\mu}_1u_1 +\tilde{\mu}_2u_2-\tilde f$.
The generalization for cases involving crystalline phases is provided in section~S1 of the SM.

Resulting phase diagrams are given in Fig.~\ref{fig:phase_diagrams}. The reciprocal reference case of Fig.~\ref{fig:phase_diagrams}(a) features four uniform phases (with high/low densities $u_i$ as indicated by the arrows in the four corners) and allows for five two- and two three-phase coexistencies. Nonreciprocity results in various changes in the phase behavior [Figs.~\ref{fig:phase_diagrams}(b)-(d)]: 
Because $\alpha \neq 0$, Fig.~\ref{fig:phase_diagrams}(b)-(d) do not show the field exchange symmetry ($u_1\leftrightarrow u_2$) of Fig.~\ref{fig:phase_diagrams}(a), however, field inversion [$(u_1,u_2)\leftrightarrow (-u_1,-u_2)$] is retained. Only the outer regions where at least one $|\bar{u}_i|$ is large, are qualitatively as in Fig.~\ref{fig:phase_diagrams}(a) as the reciprocal nonlinear parts of the $f_i$ dominate. Changes due to nonreciprocity are strongest in the central region of low $|\bar{u}_i|$. In Fig.~\ref{fig:phase_diagrams}(b) ($\xi=0.69$) the three-phase coexistence has become unstable as the coexisting phase of lowest $|\bar{u}_i|$ has crossed the threshold of the oscillatory instability. Further, the stable coexistence of $\downarrow\downarrow$- and $\uparrow\uparrow$-phase has disappeared.  Example bifurcation structures and density profiles along the highlighted red tie lines in Figs.~\ref{fig:phase_diagrams}(a) and (b) are discussed in sections S3 and S4 of the SM, respectively.

\begin{figure*}[t]
\includegraphics[width=\textwidth]{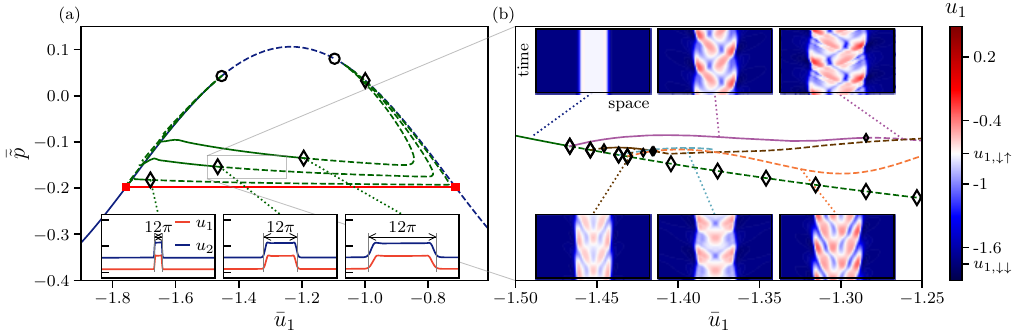}
\caption{(a) Bifurcation diagram giving the (mean) pressure $\bar{\tilde p}$ over $\bar{u}_1$ along the red tie line in the phase diagram Fig.~\ref{fig:phase_diagrams}(c) [$\bar{u}_2=1.116\bar{u}_1+1.081$]. Branches of uniform and phase-separated states are given as blue and green lines, respectively. For the latter, curves are given for three domain sizes $L=20\pi, 40\pi$ and $160\pi$ (top to bottom). The red line and squares indicate the Maxwell construction in the thermodynamic limit. Insets give the profiles at the Hopf bifurcations on the three branches.  (b) Magnification of the branch of stationary phase-separated states for $L=40\pi$ including more Hopf bifurcations than (a) and the first emerging branches of time-periodic states. The insets show space-time plots of different states where patches of oscillatory states coexist with the uniform steady background state. Solid [dashed] lines indicate linearly stable [unstable] states. Circles [diamonds] mark Cahn-Hilliard [conserved-Hopf] instabilities. Small diamonds in (b) indicate torus bifurcations, i.e. Hopf bifurcations of oscillatory states.
}
\label{fig:bif_hopf}
\end{figure*}

At larger nonreciprocity [$\xi=3$, Fig.~\ref{fig:phase_diagrams}(c)], the four pairs of binodals have entirely separated, triple point regions and $\downarrow\downarrow$-$\uparrow\uparrow$ coexistence have disappeared, and the Hopf threshold now forms a closed curve that, most importantly, intersects some binodals. In other words, part of the coexisting states are oscillatory unstable thereby predicting the coexistence of a uniform state and an oscillatory state of different mean densities. This effectively predicts the behavior found in Fig.~\ref{fig:2dcoexistence}(a,b) and can be further appreciated in the dramatically changed bifurcation structure, see Fig.~\ref{fig:bif_hopf}~(a). 
There, branches of steady states are analyzed as one follows the red tie line in Fig.~\ref{fig:phase_diagrams}(c) where the binodal at small $|\bar{u}_i|$ corresponds to a Hopf-unstable $\downarrow\uparrow$-state. At the spinodals, a branch of steady phase-separated states (green, shown for three domain sizes $L$) emerges from the branch of uniform states (blue), undergoes saddle-node bifurcations (on the left and on the right). For large domains the almost horizontal central part approaches the red Maxwell line. However, in stark contrast to Figs.~S2 and S3 of the SM, here, when increasing $\bar{u}_1$ the steady phase-separated state becomes oscillatory unstable at a Hopf bifurcation. For all considered $L$ this occurs when the unstable $\downarrow\uparrow$-phase has grown to a critical size $\ell_\mathrm{c}\approx 12 \pi$ [insets of Fig.~\ref{fig:bif_hopf}(a)] \footnote{The observed $\ell_\mathrm{c}$ is approximately four times larger than the smallest unstable wavelength, that results from the linear stability analysis of a uniform state of the $\downarrow\uparrow$-phase. However, here we observe that the oscillation extends across the phase boundary into the linearly stable $\downarrow\downarrow$-phase, where it propagates as a traveling wave with exponentially damped amplitude. Due to this damping, the first self-sustained oscillation occurs at larger domains and with larger critical wavelengths compared to the standing waves emerging from an isolated uniform $\downarrow\uparrow$-state with, e.g., Neumann boundary condition}.

 The magnification in Fig.~\ref{fig:bif_hopf}(b) shows that more such bifurcations follow and result in the emergence of several branches of time-periodic states. Intriguingly, these indeed correspond to the stable coexistence of a uniform $\downarrow\downarrow$-phase with an oscillatory $\downarrow\uparrow$-phase with domain sizes defined by the lever rule (branch in Fig.~\ref{fig:bif_hopf}(b) that bifurcates supercritically at $\bar{u}_1\approx-1.47$, and eventually destabilizes at a torus bifurcation). Corresponding time-simulations in 1d (see section~S4 of the SM) confirm the prediction of Fig.~\ref{fig:bif_hopf}. The robustness of the predicted uniform-oscillatory coexistence is further evidenced by the 2d case already presented in Fig.~\ref{fig:2dcoexistence}(a). Interestingly, large domains of the irregular oscillatory state tend to split (in 1d) or develop inner holes filled by the uniform phase (in 2d). Although, this might seem similar to the \enquote{bubbly phase separation} observed for AMB$+$ in 2d \cite{TjNC2018prx}, close inspection reveals that here the holes evolve very slowly, and do practically not fuse with each other or with the domain interface \footnote{Further, in contrast to \cite{TjNC2018prx} no additional noise is needed to create or sustain them.}. If the oscillatory domain is sufficiently small, three-, four- or five-fold waves are observed, the latter circling a periodically appearing self-organized ring structure around a central hole [Fig.~\ref{fig:2dcoexistence}(b), Movie 1(b) in SM]. Furthermore, Fig.~\ref{fig:phase_diagrams}(b) even predicts three-phase coexistence of two stable uniform phases with a weakly unstable oscillatory one. This is indeed found in Fig.~\ref{fig:2dcoexistence}(c) where the blue unstable phase shows irregular bursts of waves interspersed with long \enquote{calm} phases.

\begin{figure}[tb]
\centering
\includegraphics[width=\columnwidth]{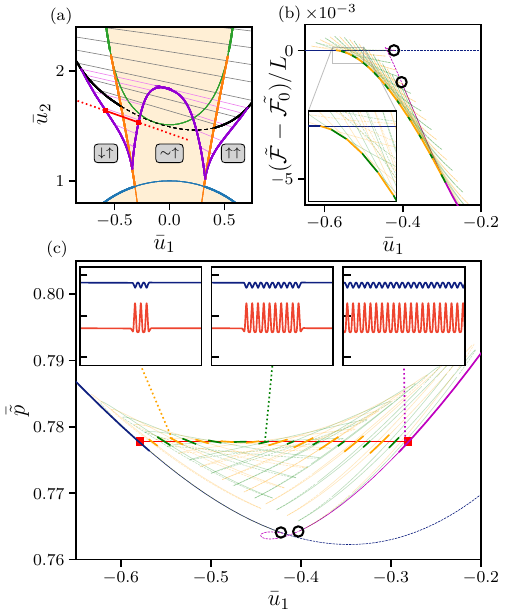}
\caption{Panel (a) magnifies part of Fig.~\ref{fig:phase_diagrams}(d) focusing on the coexistence region of uniform and crystalline phases. Panels (b) and (c) analyze the bifurcation structure along the red tie line in (a)[$\bar{u}_2=-0.353\,\bar{u}_1+1.431$] providing mean energy and pressure $\bar{\tilde p}$ over $\bar{u}_1$, respectively, Shown are uniform (blue), periodic (magenta), and even [odd] (green [yellow]) localized states, i.e., crystallites. Panels (b) and (c) highlight the states of lowest energy as piecewise thick lines, thereby in (c) illustrating how the Maxwell line (red horizontal) is approached. Parameters are as in Fig.~\ref{fig:bif_turing}~(d)  and $L=80\pi$.}
\label{fig:bif_turing}
\end{figure}

Focusing next on the phase diagram of Fig.~\ref{fig:phase_diagrams}(d) we see that further increasing the nonreciprocity $\xi$ separates the binodals from the Hopf-unstable region. In contrast to the previous cases, Fig.~\ref{fig:phase_diagrams}(d) features a conserved-Turing instability (case of nonunity rigidity ratio, $\kappa \neq 1$) giving rise to crystalline phases. The resulting (purple) binodals represent their coexistence with uniform phases, see magnification in Fig.~\ref{fig:bif_turing}(a). With increasing $|\bar u_2|$, the lattice spacing of the crystalline states increases and finally diverges at $(\bar{u}_1,\bar{u}_2)\approx(0,1.9)$ (not shown). In other words, the crystalline phase transforms into a phase-separated state with a smooth transition between the binodals. 
In contrast, when decreasing $|\bar u_2|$ the coexistence ranges shrink till they terminate in tricritical points where the phase transition changes from first to second order. Figs.~\ref{fig:bif_turing}(b) and~\ref{fig:bif_turing}(c) show the bifurcation structure along the red tie line in Fig.~\ref{fig:bif_turing}(a) in terms of mean energy and pressure, respectively.  Where the uniform state (stable at small $\bar{u}_1$) undergoes the conserved-Turing instability, a branch of periodic (crystalline) states emerges, itself soon spawning two branches of localized states forming the typical \enquote{snakes and ladders} structure \cite{BuKn2006pre,Knob2016ijam,HAGK2021ijam}.
The ensuing multistability represents phase coexistence: The heavy lines in Figs.~\ref{fig:bif_turing}(b,c) highlight respective states of minimal spurious energy $\tilde{\mathcal{F}}$. 
Also see the example of a 2d crystallite in Fig.~\ref{fig:2dcoexistence}(d). In this way, Fig.~\ref{fig:bif_turing}(c) reveals that the resulting piecewise curve defines a narrow horizontal band centered about the coexistence pressure. With increasing domain size the band will become thinner and approach the Maxwell line as known from the passive PFC model~\cite{TFEK2019njp}. The validity of the approach of \cite{TFEK2019njp} for the studied NRCH model impressively evidences the power of the spurious gradient dynamics form.

To conclude, we have shown that several aspects of the equilibrium thermodynamics of phase transitions can be applied to a third class of models between passive and active systems, that features a spurious gradient dynamics structure. In particular, employing a specific NRCH model we have illustrated that the resulting (spurious) Maxwell construction allows one to predict its phase behavior, intriguingly, including the nonequilibrium two- and three-phase coexistence of uniform and oscillatory states featuring large-amplitude regular or irregular waves. We expect such coexistencies to widely occur in systems with conservation laws that show large-scale oscillatory instabilities \cite{FrTP2023ptrsa,FrTh2023prl,GrTh2024c}, see e.g., \cite{PaPa2023prr,DAGM2023prl}. Notably, this includes crystalline (or micro-phase separated) states that also only emerge due to nonreciprocity. Extending the calculation of spurious pressure and chemical potential to crystalline phases (see SM) we have further shown that the full characterization of first order phase transitions via bifurcation diagrams developed for PFC models \cite{TFEK2019njp} applies in the present nonequilibrium case.
  
Our analysis of the specific NRCH model already indicates the validity of the approach for all models in the class of spurious gradient dynamics, like active PFC models \cite{MeLo2013prl,OpGT2018pre,VHKW2022msmse}, coupled Cahn-Hilliard and Swift-Hohenberg models \cite{FHKG2023pre}, and the FitzHugh-Nagumo RD model \cite{ScBP1995pd}\footnote{Here, the list comprises models with purely mass-conserving, purely nonmass-conserving and mixed dynamics. Note that further generalizations of the spurious gradient structure will surely exist as other similar structures have been described that only partially overlap with the form introduced in \cite{FHKG2023pre}. This includes the skew-gradient dissipative systems of \cite{KuYa2003pd}, cf.~discussion in conclusion of \cite{FHKG2023pre}, the Maxwell-like construction in \cite{SSCK2018pre}, see section~S5 of SM, and the mechanistic discussion of coexistence in \cite{ORMB2023pnasusa}.}.
All of these could be (re-)investigated defining corresponding nonequilibrium chemical potentials $\tilde\mu_i$ and pressures $\tilde p$ as described in the SM which, in turn, can be combined with extensions of mechanistic views of microscopic systems\cite{ORMB2023pnasusa,IBHD2015PRX}. Resulting nonequilibrium multispecies coexistence conditions, e.g., \cite{ChEO2024preprint,Dinelli2024Paris}, may then be further unified and employed to further elucidate the interplay of conservation laws and nonreciprocity.

\end{document}


\title{Supplementary material -- Coexistence of uniform and oscillatory states resulting from nonreciprocity and conservation laws}

\author{Daniel Greve}  
\email{daniel.greve@uni-muenster.de}
\affiliation{Institut für Theoretische Physik, Universität Münster, Wilhelm-Klemm-Str. 9, 48149 Münster, Germany}

\author{Giorgio Lovato}  
\email{stu242274@mail.uni-kiel.de}
\affiliation{Institut für Theoretische Physik und Astrophysik, Christian-Albrechts-Universität zu Kiel, Leibnizstr. 5, 24098 Kiel,Germany}

\author{Tobias Frohoff-Hülsmann}
\email{t\_froh01@uni-muenster.de}  
\affiliation{Institut für Theoretische Physik, Universität Münster, Wilhelm-Klemm-Str. 9, 48149 Münster, Germany}

\author{Uwe Thiele}  
\email{u.thiele@uni-muenster.de}
\homepage{www.uwethiele.de}
\affiliation{Institut für Theoretische Physik, Universität Münster, Wilhelm-Klemm-Str. 9, 48149 Münster, Germany}
\affiliation{Center for Nonlinear Science (CeNoS), Universität Münster, Corrensstr. 2, 48149 Münster, Germany}
\affiliation{Center for Multiscale Theory and Computation (CMTC), Universität Münster, Corrensstr. 40, 48149 Münster,Germany}

\begin{abstract}
Here, we provide additional information to accompany the main text: First, we describe the theoretical background for the construction of the phase diagrams, i.e., for the  calculation of binodals and spinodals. To allow for a deeper understanding of the phase behavior in finite systems, we also provide bifurcation diagrams for the passive case as well as for weakly active cases. For the latter we discuss the occurrence of interfaces with oscillatory tails. We then briefly discuss the relation of the employed spurious Maxwell construction to related approaches, e.g., as described for active model B. Finally, we give details on the employed numerical methods and parameters.
\end{abstract}

\maketitle

\tableofcontents
\newpage
\section{Spurious Maxwell construction including crystalline phases}\label{sec:MW_NRCH}
Here, we explain how the spurious Maxwell construction arises from the underlying spurious gradient dynamics structure employing the linearly coupled nonreciprocal Cahn-Hilliard (NRCH) model as example. Note that the concept of spurious gradient dynamics was established in Ref.~\cite{FHKG2023pre} where, however, the entire discussion focused on the explanation of the unexpected existence of resting asymmetric (\enquote{broken-parity}) states numerically observed for a diverse set of nonvariational models with purely mass-conserving, purely non-mass-conserving and mixed dynamics. Before, it was assumed that all asymmetric states in nonvariational models must move. The detailed mathematical argument of Ref.~\cite{FHKG2023pre} showed that asymmetric states may indeed remain at rest if the model is of a certain special form, termed  \enquote{spurious gradient dynamics.}  Ref.~\cite{FHKG2023pre} analyzed a number and variety of examples and explained why the bifurcation diagrams for all these models feature branches of resting asymmetric states. However, further physical consequences, in particular, regarding the nonequilibrium phase behavior were not considered. The present work discusses the spurious Maxwell construction as a further far-reaching  implication of such a structure that in combination with linear stability analyses allows one to quite precisely predict a range of intricate nonequilibrium coexistences of two or more uniform and oscillatory phases. This is only possible for two or more species.

The specific  NRCH model used is
\begin{equation}\label{eq:nrch}
\partial_tu_i=\vec\nabla \cdot \left[Q_i\vec \nabla \left(\frac{\delta\mathcal{F}}{\delta u_i} + \mu_{i}^\mathrm{nr}\right)\right].
\end{equation}
with 
$\mu_{1}^\mathrm{nr}=-\alpha u_2$, $\mu_{2}^\mathrm{nr}=+\alpha u_1$ and $Q_1=Q_2=1$. The underlying energy is $\mathcal{F}=\mathcal{F}_1[u_1]+\mathcal{F}_2[u_2]+\mathcal{F}_{12}[u_1,u_2]$, where $\quad \mathcal{F}_i [u_i]= \int_\Omega \left[ \frac{\kappa_i}{2} |\vec{\nabla} u_i|^2 + f_i(u_i)\right] \mathrm{d}^n x$, $f_i(u_i)=au_i+u_i^3$, $\kappa_1=1$, $\kappa_2=\kappa$ and $\mathcal{F}_\mathrm{12}=\int_\Omega f_{12}(u_1,u_2) \mathrm{d}^n x$ with $f_{12}=-\rho u_1 u_2$. Without nonequilibrium chemical potentials $\mu_{i}^\mathrm{nr}$, i.e., for $\alpha=0$ Eq.~\eqref{eq:nrch} represents a proper gradient dynamics that describes the temporal evolution of the two scalar fields $\vecg u=(u_1(\vec x,t),u_2(\vec{x},t))$, i.e., the $M_i$ are positive and $\mathcal{F}$ is a proper Lyapunov functional.

However, as discussed in the main text, even for $\alpha\neq0$ it can be formally written in gradient dynamics form:
\begin{equation}\label{eq:gradDyn}
  \partial_t \vecg{u}=\vec \nabla\cdot\left[
    \tilde{\tensg{M}}\vec \nabla\frac{\delta \tilde{\mathcal{F}}}{\delta\vecg{u}}
    \right],
\end{equation}
with the transformed functional
\begin{equation}\label{eq:ps_Max_NRCH}
\begin{split}
	 \tilde{\mathcal{F}}=\frac{\rho}{\rho+\alpha}\mathcal F_1+\frac{\rho}{\rho-\alpha}\mathcal F_2+\mathcal{F}_\mathrm{12} \quad \text{and mobility matrix} \quad
	\quad\tensg{\tilde M}=\begin{pmatrix}\frac{\rho+\alpha}{\rho}&0\\0&\frac{\rho-\alpha}{\rho}\end{pmatrix}.
\end{split}
\end{equation}
In contrast to the passive case ($\alpha=0$) the gradient dynamics is now \enquote{spurious} as for dominant nonreciprocity $\zeta=\alpha^2-\rho^2>0$, the mobility matrix is not positive definite and the functional $\tilde{\mathcal{F}}$ is not bounded from below. However, in the analysis of the model we can formally proceed as for a proper gradient dynamics. To clearly mark all quantities and procedures determined on this basis we will accordingly  call them \enquote{spurious}, e.g., spurious pressure or spurious Maxwell construction.

We set $\partial_t \vecg{u}=0$ in Eq.~\eqref{eq:gradDyn}, integrate, and set the first integration constant to zero to impose the condition of zero net flow across the
boundaries (e.g., realized by Neumann or periodic boundary conditions). We multiply with $\tilde{\tensg{M}}^{-1}$ and integrate again to obtain
\begin{equation}\label{eq:chem_pot}
	\frac{\delta \tilde{\mathcal{F}}}{\delta\vecg{u}}=\tilde{\vecg \mu}=\text{const.}
\end{equation}
Hence, the steady state equations \eqref{eq:chem_pot} can be seen as arising from the functional variation of a spurious grand potential $\tilde{\Omega}=\tilde{\mathcal{F}}-\int_\Omega \vecg u \tilde{\vecg \mu}\mathrm{d}^n x=\int_\Omega \tilde{\omega} \mathrm{d}^n x$, with the spurious grand potential density 
\begin{equation}\label{eq:grand_potential}
\tilde{\omega} = \frac{\rho}{\rho+\alpha}\left[f_1(u_1)+\frac{1}{2}|\vec \nabla u_1|^2\right]+\frac{\rho}{\rho-\alpha}\left[f_2(u_2)+\frac{\kappa}{2}|\vec \nabla u_2|^2\right]+f_{12}(u_1,u_2)-u_1\tilde{\mu}_1-u_2\tilde{\mu}_2. 
\end{equation}
Note that for heterogeneous (e.g., crystalline) states $\tilde{\omega}$ is normally not uniform in space but in analogy to analytical mechanics may be seen as a spatial Lagrangian. Then, we can determine the corresponding spatial Hamiltonian $\tilde{\mathcal{H}}$ that is indeed uniform in space. To do so, we identify the $u_i$ as generalized positions and define the appropriate generalized momenta as $\vec{\pi}_i=\frac{\partial \tilde{\omega}}{\partial (\vec \nabla u_i)}$ to obtain
\begin{equation}\label{eq:hamiltonian}
	\tilde{\mathcal{H}}=\tilde{\omega}-\sum_{i=1}^2\vec{\pi}_i\cdot\vec \nabla u_i= \frac{\rho}{\rho+\alpha}\left[f_1(u_1)-\frac{1}{2}|\vec \nabla u_1|^2\right]+\frac{\rho}{\rho-\alpha}\left[f_2(u_2)-\frac{\kappa}{2}|\vec \nabla u_2|^2\right]+f_{12}-u_1\tilde{\mu}_1-u_2\tilde{\mu}_2. 
\end{equation}
$\tilde{\mathcal{H}}$ is a first integral of the steady state equation \eqref{eq:chem_pot}, i.e. a spatially conserved quantity fulfilling $\vec \nabla \tilde{\mathcal{H}}=\vec{0}$. Hence, in the context of phase coexistence the value of $\tilde{\mathcal{H}}$ is equal for all coexisting phases and even across all interfaces.  For coexisting uniform states, the spatial derivatives within $\tilde{\mathcal{H}}$ are zero in the respective phases resulting in the spurious pressure 
\begin{equation} \label{eq:pressure}
	\tilde{p}=	u_1\tilde{\mu}_1+u_2\tilde{\mu}_2-\tilde{f},
\end{equation}
where $\tilde f=\frac{\rho}{\rho+\alpha}f_1+\frac{\rho}{\rho-\alpha}f_2+f_{12}$ and the spurious chemical potential reduces to $\tilde{\mu}_i=\partial \tilde{f}/\partial u_i$. 
Hence, the combination of Eqs.~\eqref{eq:chem_pot} and  \eqref{eq:pressure} gives the binodals for two coexisting uniform phases A and B. Three of the values of the coexisting concentrations are determined by  
\begin{align}\label{eq:coexistence_cond_NRCH}
    \tilde{\mu}_1(\vecg u^{(A)})&=\tilde{\mu}_1(\vecg u^{(B)}),\quad \tilde{\mu}_2(\vecg u^{(A)})=\tilde{\mu}_2(\vecg u^{(B)})\nonumber\\
     \quad \text{and} \quad \tilde p(\vecg u^{(A)})&=\tilde p(\vecg u^{(B)})
\end{align}
while the fourth provides a degree of freedom, i.e., binodals are pairs of lines in the $(\bar u_1,\bar u_2)$-plane, see e.g., Fig.~2 of the main text. The described construction generalizes the standard Maxwell construction of equilibrium thermodynamics to coexisting uniform steady states in spurious gradient dynamics, i.e., it provides a \enquote{spurious Maxwell construction}. The resulting binodals are obtained by numerical continuation \cite{HoAT2021jpcm}.

Next, we extend the Maxwell construction to crystalline phases, i.e., for cases where a spatially periodic phase coexists with a uniform phase. Here we focus on one spatial dimension where the lattice constant of the crystal simply corresponds to a spatial period $L$.

In the thermodynamic limit, the periodic phase assumes the value of $L$ of extremal grand potential for one period at fixed chemical potential, i.e. $(\frac{\partial}{\partial L} \tilde{\Omega}/L)_{\vecg \mu=\text{const.}}=0$ (or equivalently of extremal free energy for one period at fixed mean particle density $(\frac{\partial}{\partial L} \tilde{\mathcal F}/L)_{\bar{\vecg u}=\text{const.}}=0$). Therefore, the relevant $L$ is obtained using the additional condition\footnote{This nontrivial identity can be obtained as follows. We introduce a length scale $\hat{x}=x/L$ to rescale one period to unit length. In terms of the new scales, we then have 
\begin{equation*}
\begin{split}
\frac{\tilde{\Omega}[\vecg u(L)]}{L}&=\hat{\tilde{\Omega}}[\vecg u(L),L]=\int_0^1 \hat{\tilde{\omega}} \mathrm{d}\hat{x}\\
\text{with} \qquad \hat{\tilde{\omega}} &= \frac{\rho}{\rho+\alpha}\left[f_1(u_1)+\frac{1}{2}\left|\frac{1}{L}\partial_{\hat{x}} u_1\right|^2\right]+\frac{\rho}{\rho-\alpha}\left[f_2(u_2)+\frac{\kappa}{2}\left|\frac{1}{L}\partial_{\hat{x}} u_2\right|^2\right]+f_{12}-u_1\tilde{\mu}_1-u_2\tilde{\mu}_2. 
\end{split}
\end{equation*}
Note, that the dependence of $\hat{\tilde{\Omega}}$ on $L$ is both explicit and implicit as the fields $\vecg u$ depend on $L$. This is due to the solution of a partial differential equation (PDE) depending on the domain extend, i.e., the shape of the profile depends on the length of a period. The PDE rescales as
\begin{equation*} \label{eq:func_deriv_rescaled}
\frac{\delta \tilde{\Omega}}{\delta \vecg u}=0 \quad \text{with periodic BC on [0,L]}, \qquad \implies \qquad 
\frac{\delta \hat{\tilde{\Omega}}}{\delta \vecg u}=0 \quad \text{with periodic BC on [0,1]},
\end{equation*}
where the period $L$ no longer occurs in the BC but in the derivatives $\partial_x \rightarrow \frac{1}{L}\partial_{\hat{x}}$. We can then evaluate the derivative using the functional chain rule for the implicit derivative
\begin{equation*}
\begin{split}
\left(\frac{\partial }{\partial L} \frac{\tilde{\Omega}[\vecg u(L)]}{L}\right)_{\vecg \mu=\text{const.}}&=\left(\frac{\partial }{\partial L} \hat{\tilde{\Omega}}[\vecg u(L),L]\right)_{\vecg \mu=\text{const.}}=\int_0^1 \underbrace{\frac{\delta \hat{\tilde{\Omega}} }{\delta \vecg u}}_{=0}\frac{\partial \vecg u}{\partial L}\mathrm{d}\hat{x}+ \int_0^1 \frac{\partial \hat{\tilde{\omega}}}{\partial L}  \mathrm{d}\tilde{x}\\
&=-\frac{2}{L}\int_0^1 \left[\frac{\rho}{2(\rho+\alpha)}\left|\frac{1}{L}\partial_{\hat{x}} u_1\right|^2+\frac{\kappa \rho}{2(\rho-\alpha)} \left|\frac{1}{L}\partial_{\hat{x}} u_2\right|^2 \right]\mathrm{d} \hat{x}\\
&=-\frac{2}{L^2}\int_0^L \left[\frac{\rho}{2(\rho+\alpha)}|\partial_x u_1|^2+\frac{\kappa \rho}{2(\rho-\alpha)} |\partial_x u_2|^2 \right]\mathrm{d} x,
\end{split}
\end{equation*}
where in the last step we reintroduced the original lengthscale $x$. 
}

\begin{equation}\label{eq:wavelength_constraint}
\begin{split}
	\left(\frac{\partial }{\partial L} \frac{\tilde{\Omega}[\vecg u(L)]}{L}\right)_{\vecg \mu=\text{const.}}&=-\frac{2}{L^2}\int_0^L \left[\frac{\rho}{2(\rho+\alpha)}|\partial_x u_1|^2+\frac{\kappa \rho}{2(\rho-\alpha)} |\partial_x u_2|^2 \right]\mathrm{d} x=0 
\end{split}.
\end{equation}
To obtain the thermodynamic quantities necessary to construct the binodals, we numerically solve Eqs.~\eqref{eq:chem_pot} in a domain corresponding to one period $L$ of the periodic phase (see Ref.~\cite{HoAT2021jpcm}), where we allow for $L$ to adjust via enforcement of Eq.~\eqref{eq:wavelength_constraint} as a side condition.

Further, note that averaging Eqs.~\eqref{eq:grand_potential}, \eqref{eq:pressure} over one period and adding multiples of Eq.~\eqref{eq:wavelength_constraint} yields $-\bar{\tilde{p}}=\bar{\tilde{\omega}}=\tilde{\mathcal{H}}$ for the periodic phase, i.e., although the spurious pressure and grand potential density differ and vary within one period of the crystalline phase, their spatial averages are identical and can -- equivalently to the spatial Hamiltonian -- be used as a coexistence condition, i.e. Eqs.~\eqref{eq:coexistence_cond_NRCH} still hold for periodic phases  when $\tilde{p}$ is replaced by its spatial average $\bar{\tilde{p}}$. 

\newpage
\section{Spinodals -- linear stability analysis}\label{sec:lin_stab}

\begin{figure}
\centering
\includegraphics[width=.7\textwidth]{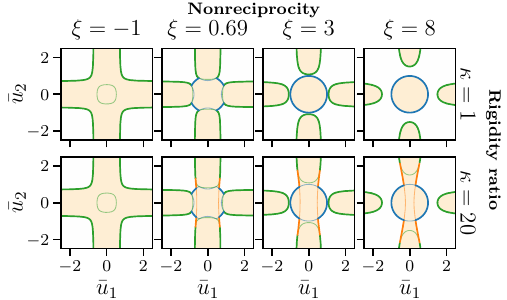}
\caption{Linear stability thresholds for uniform steady states in the plane of mean concentrations for the NRCH model for increasing nonreciprocity parameter $\xi=\alpha^2-\rho^2$ (left to right) as well as unity ($\kappa=1$, top row) and nonunity ($\kappa=20$) rigidity ratio, i.e., without and with a conserved-Turing instability, respectively. Further, $a=-1.5$. Instability is indicated by orange shading. Green, orange and blue thick [thin] lines indicate primary [secondary] Cahn-Hilliard, conserved-Turing and conserved-Hopf instabilities, respectively.}
\label{fig:spinodals_CH}
\end{figure}

Since the NRCH model describes two conserved quantities any homogeneous state $(\bar u_1, \bar u_2)$ is steady, i.e., time independent. However, their linear stability  with respect to small perturbations may differ. To determine the instability thresholds and their dependence on the rigidity ratio $\kappa$ and the nonreciprocity parameter $\xi=\alpha^2-\rho^2$ we perform a linear stability analysis. 
Thereby, we adapt the analyses presented in Refs.~\cite{FrWT2021pre,FrTP2023ptrsa}.
Using the ansatz $(u_1,u_2)=(\bar u_1,\bar u_2) + \varepsilon (\hat u_1,\hat u_2)\,\exp(\lambda t+ \text{i} \vec{k}\cdot\vec{x})$ and linearizing in the smallness parameter $\varepsilon\ll1$ we obtain
\begin{equation} 
   \left(-k^2 \tens{L}(k^2)-\lambda \tens{\mathbf{1}}\right)\,\left(\begin{array}{c}\hat u_2\\ \hat u_2 \end{array} \right)=\vecg{0}
\label{91jac}
\end{equation} 
with the reduced Jacobi matrix\footnote{Note that, in a more general case with a non-identity mobility matrix $\tens M$, which can possibly also depend on $\vecg u$, the linearized equation instead reduces to\begin{equation*} 
   \left(-k^2 \tens M_0\tens{L}(k^2)-\lambda \tens{\mathbf{1}}\right)\,\left(\begin{array}{c}\hat u_2\\ \hat u_2 \end{array} \right)=\vecg{0},
\end{equation*} where $\tens M_0=\tens M(\bar{u}_1,\bar{u}_2)$. This may alter the thresholds for conserved-Hopf and conserved-Turing instabilities whereas Cahn-Hilliard thresholds remain unchanged.}
\begin{equation}
    \tens{L}(k)= \left( \begin{array}{cc} k^2 + a + 3 \bar u_1^2 & -(\rho+\alpha) \\  -(\rho-\alpha) &   \kappa k^2 + a + 3 \bar u_2^2
  \end{array} \right)\,.
\end{equation}

Solving the linear system, the two eigenvalues are 
\begin{align}
\lambda_\pm(k)=-k^2\tilde{\lambda}_{\pm}(k)= &-\frac{k^2}{2} \left(\Tr \tens{L}(k^2) \pm \sqrt{\left(\Tr \tens{L}(k)\right)^2 - 4 \Det \tens{L}(k^2)}\right), ~\label{eq:evs_NRCH}\\
\text{with} \quad \Tr \tens{L}(k^2)&=2a+3(\bar{u}_1^2+\bar{u}_2^2)+k^2(1+\kappa)\\ \text{and} \quad \Det\tens{L}(k^2)&=(a+3\bar{u}_1^2)(a+3\bar{u}_2^2)+\xi+k^2[\kappa(a+3\bar{u}_1^2)+(a+3\bar{u}_2^2)]+\kappa k^4. 
 \label{eq:lambda}  \end{align}

As further explained in Ref.~\cite{FrTP2023ptrsa} the linear stability of the NRCH model is simply the conserved pendant with an overall factor $(-k^2)$ in \eqref{eq:evs_NRCH} of the linear stability of a two-component (nonconserved) reaction-diffusion system \cite{Turi1952ptrslsbs}.  Then, the NRCH model exhibits three different instabilities~\cite{FrWT2021pre,FrTP2023ptrsa} as we briefly summarize (using the classification introduced in the Supplementary Material of Ref.~\cite{FrTh2023prl}):
\begin{itemize}
    \item[(i)] The Cahn-Hilliard instability: A conserved large-scale stationary instability that occurs when the real-valued $\tilde \lambda_+(k_\text{c}=0)$ crosses zero. Therefore, the threshold is at $\Det \tens{L}(k=0)=0$ for $\Tr\tens{L}(k=0)<0$, i.e., 
    \begin{equation}\label{eq:lin_CH}
        (a+3\bar{u}_1^2)(a+3\bar{u}_2^2)+\xi=0 \qquad \text{and}\qquad 2a+3(\bar{u}_1^2+\bar{u}_2^2)<0.
    \end{equation}

    \item[(ii)] The conserved-Hopf instability: A conserved large-scale oscillatory instability that occurs when $\Re \tilde \lambda_\pm(k=0)$ crosses zero while $\Im \tilde \lambda_\pm(k=0)\neq 0$ (see Ref.~\cite{GrTh2024c} for an extended discussion of this type of instability). Therefore, the threshold is at $\Tr \tens{L}(k=0)=0$ for $\Det \tens L(k=0) >0$, i.e., 
    \begin{equation}\label{eq:lin_Ho}
       2a+3(\bar{u}_1^2+\bar{u}_2^2)=0 \qquad \text{and} \qquad (a+3\bar{u}_1^2)(a+3\bar{u}_2^2)+\xi>0. 
    \end{equation}
        \item[(iii)] The conserved-Turing instability: A conserved small-scale stationary instability that occurs when a maximum of $\tilde{\lambda}_+$ crosses zero for a nonvanishing $k_\text{T}$, i.e., when $\tilde{\lambda}_+(k=k_\text{T})=0$ and $\partial \tilde{\lambda}_+/\partial(k^2)(k=k_\text{T})=0$. This results in  (cf.~Ref.~\cite{FrTP2023ptrsa})
        \begin{equation}\label{eq:kT}
            k_\text{T}^2= -\frac{a+3\bar{u}_2^2+\kappa(a+3\bar{u}_1^2)}{2\kappa}
        \end{equation}
        and the instability occurs when
    \begin{equation}\label{eq:lin_Tu}
        \left[\kappa(a+3\bar{u}_1^2)-(a+3\bar{u}_2^2)\right]^2-4\kappa\xi=0\, \quad \text{and} \quad k_\text{T}^2>0. 
      \end{equation}
Note that sometimes the notion \enquote{Turing instability} is used in a narrow sense for a small-scale stationary instability in a reaction-diffusion system (as caused by an activator and an inhibitor with a contrast in diffusion) or even confusingly for any stationary instability in such systems. Here we employ the notion for any \textit{small-scale} stationary instability.
\end{itemize}
These three conditions result in linear stability thresholds (spinodals) that are displayed, e.g., in the $(\bar u_1,\bar u_2)$-plane, see e.g., Fig.~\ref{fig:spinodals_CH}. Note that Fig.~2 of the main text contains the thresholds from the three panels on the left of the top row and from the final panel on the right of the bottom row of Fig.~\ref{fig:spinodals_CH}.

\section{Bifurcation analysis for the passive case}
\begin{figure}
\includegraphics[width=\textwidth]{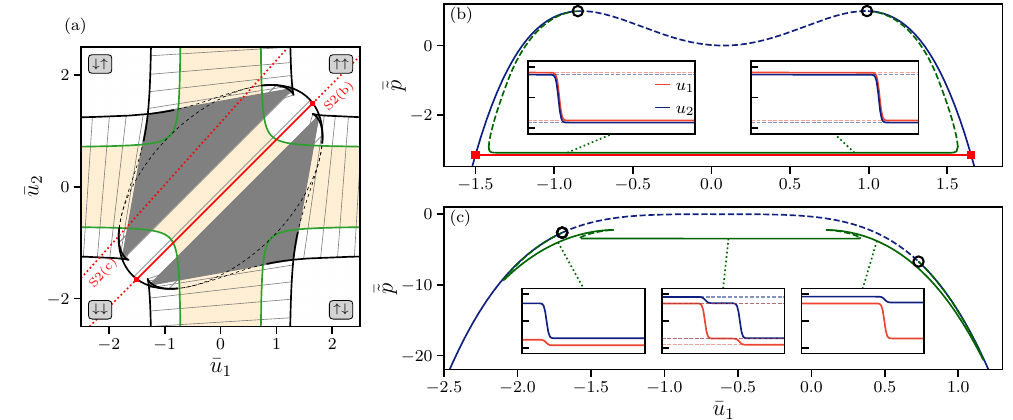}
\caption{Bifurcation diagrams in terms of the pressure $\tilde p$ as a function of $\bar{u}_1$ in the passive case $\alpha=0$ for a finite domain size ($L=20\pi$). Panels (b) and (c) show the behavior along the two tie lines [(b) $\bar u_2=\bar u_1 -0.151$ and (c) $\bar u_2=1.116\,\bar u_1 + 1.153$] marked in panel (a) corresponding to Fig.~2(a) of the main text. Blue [green] lines correspond to branches of uniform [phase-separated] states while solid [dashed] lines indicate stability [instability]. Circles mark pitchfork bifurcations, and insets show typical concentration profiles for two- and three-phase coexistence, where the dashed horizontal lines indicate the predicted plateau values, i.e., the binodal values.}
\label{fig:bif_ana_passive.pdf}
\end{figure}

To allow for a comparison to the active cases, we briefly review the passive (reciprocal) case ($\alpha=0$) that is extensively analyzed elsewhere \cite{FrWT2021pre}. It allows for five two- and two three-phase coexistences of the four uniform phases characterized by the arrows in the four corners (e.g., $\uparrow\downarrow$ indicates high $u_1$ and low $u_2$). The bifurcation structure along the two highlighted lines in Fig.~\ref{fig:bif_ana_passive.pdf}(a) is given in panels (b) and (c) in terms of the pressure $\tilde p$ and consists of branches of uniform and phase-separated states. For two-phase [three-phase] coexistence, Fig.~\ref{fig:bif_ana_passive.pdf}(b) [Fig.~\ref{fig:bif_ana_passive.pdf}(c)] follows a tie line [crosses a triple point region]. Although the considered domain is not very large (see insets), Fig.~\ref{fig:bif_ana_passive.pdf}(a) indicates how the Maxwell line (in red) is approached (cf.~\cite{TFEK2019njp}) and all insets of Fig.~\ref{fig:bif_ana_passive.pdf} show the step-like fronts between phases. Further, the plateau values along the Maxwell line and in the triple point region coincide with the predicted values from the phase diagram (bright dashed lines). Note that for the three-phase coexistence we have in total six equations for the coexisting six concentrations in phases A, B and C. This implies that there is no freedom and coexistence is restricted to triplets of points in the $(\bar u_1,\bar u_2)$-plane (corners of the gray triangles in Fig.~\ref{fig:bif_ana_passive.pdf}(a)).

\section{Spatial eigenvalues and oscillatory tails}
 Similar to the (temporal) eigenvalues (EVs) that have been calculated in the linear stability analysis in section~\ref{sec:lin_stab}, on may determine the spatial EVs. They describe how plateau values of steady states are approached in a small perturbation regime. Here, the ansatz is
\begin{equation}
\vecg u(x) = \vecg{\bar u} + \vecg{\hat u} e^{\beta  x}
\end{equation}
where $\beta$ is the spatial EV.  Linearizing in the perturbation $\vecg{\hat u}$ and solving for EVs $\beta$ gives
\begin{equation}
\beta_{1,2,3,4}=\pm\frac{1}{\sqrt{2\kappa}}\sqrt{a(1+\kappa)+3(\kappa\bar{u}_1^2+\bar{u}_2^2)\pm\sqrt{\left[a(1-\kappa)+3(\kappa\bar{u}_1^2- u_2^2)\right]^2-4\kappa\xi}}\,.
\end{equation}
The four spatial EVs describe the leading edge behavior of the \enquote{fronts} that represent the interfaces between uniform phases. For two-phase coexistence the profile may approach the two plateau values monotonically (real $\beta$) or oscillatory (complex $\beta$). For parity-symmetric systems as the present one, $\beta_1=-\beta_2$ and $\beta_3=-\beta_4$ always hold. In the dominantly reciprocal case ($\xi<0$), the  inner root is always positive, spatial EVs with nonzero real and imaginary parts can not occur. 
Thus, all phase-separated states  can only exhibit monotonic exponential approaches to the plateau value as known from the passive Cahn-Hilliard case. However, in the dominantly nonreciprocal case ($\xi>0$) complex EVs with nonzero real and imaginary parts can occur, i.e., plateaus may be approached in an oscillatory manner.  

\begin{figure*}[h]
\includegraphics[width=\textwidth]{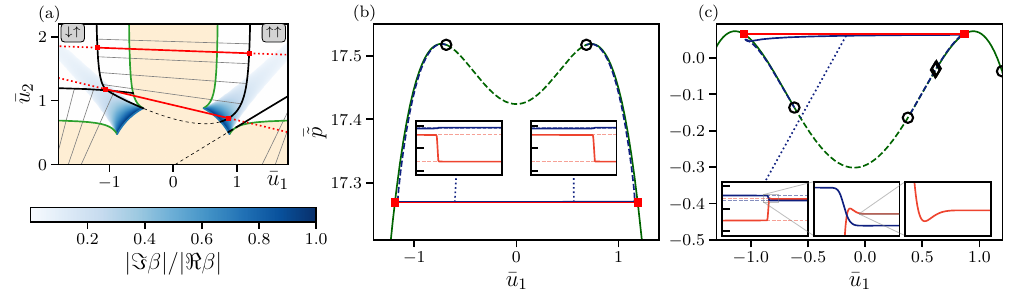}
\caption{Panel (a) magnifies part of Fig.~2(b) of the main text, i.e., the weakly active case at $\xi=0.69$ while (b) and (c) give the bifurcation diagrams ($\tilde p$ over $\bar{u}_1$) along the red tie lines in (a) for $L=60\pi$, i.e., (b) $\bar u_2(\bar u_1) =-0.037 \,\bar u_1 +1.785$ and (c) $\bar u_2(\bar u_1) =-0.236 \,\bar u_1 +0.929$. The blue shading in (a) encodes the ratio $|\Im\beta /\Re \beta|$ for spatial EVs of stable uniform states. Dashed horizontal lines in the in the insets in panels (b) and (c) indicate the predicted plateau values, i.e., the binodals.}
\label{fig:bif_tails}
\end{figure*}

The occurrence of complex EV is indicated in Fig.~\ref{fig:bif_tails} (a magnified part of Fig.~2(b) of the main text) by blue shading whose strength indicates the relative importance of the imaginary part as compared to the real part. Then, the bifurcation diagrams Fig.~\ref{fig:bif_tails}~(b) and (c) correspond to cuts through the ($\bar u_1,\bar u_2$)-plane along the tie lines highlighted in Fig.~\ref{fig:bif_tails}~(a). First, they clearly confirm the validity of the spurious Maxwell construction: For large systems, coexisting concentrations approach the predicted binodal values, the relative domain extensions are governed by the lever rule and the branch of stable phase-separated states indeed approaches the horizontal Maxwell line of constant pressure as well known from the passive case \cite{TFEK2019njp}.

Nevertheless, Fig.~\ref{fig:bif_tails}~(c) reveals that in contrast to passive phase separation the profile can be nonmonotonic, i.e., starting on the low-concentration plateau on the left, the red profile overshoots the concentration value of the upper plateau and approaches it in a (strongly damped) spatial oscillation, consistent with the occurrence of complex spatial EVs. 

Their relevance is shown in Fig.~\ref{fig:bif_tails}~(a) by the blue shading that indicates the ratio of imaginary part and real part of the spatial EVs. In the regions of darker blue the ratio becomes large and the oscillatory tails are more prominent as confirmed by the insets in Fig.~\ref{fig:bif_tails}~(c): As the right [left] plateau value corresponds to a dark [light] blue region in  Fig.~\ref{fig:bif_tails}~(a) the right [left] plateau is approached with a  more [less] pronounced spatial oscillation. The overshoot for the right plateau is of the order of $10^{-1}$ and well visible in the magnifying insets while the one for the left plateau is much smaller (of order $10^{-5}$) and not visible at the same magnification. Such an overshooting can have far reaching consequences and ultimately result in the suppression of coarsening as the interaction between neighboring interfaces changes nonmonotonically (not shown).

\section{Spurious gradient dynamics and active model B}
As pointed out in the main text and above in section~\ref{sec:MW_NRCH} the here studied two-field NRCH model corresponds to a specific example of the general spurious gradient dynamics established in \cite{FHKG2023pre} for an arbitrary number of scalar fields with either purely mass-conserving, purely non-mass-conserving or mixed dynamics. In consequence, the here discussed spurious Maxwell construction and resulting phase coexistencies can equally be studied for the entire class of multi-field spurious gradient dynamics models.

Noting the parallels to the Maxwell construction for the one-field active model~B (AMB) discussed in Refs.~\cite{WTSA2014nc,SSCK2018pre}, here, we furthermore show that both are limiting cases of a more general multi-species spurious gradient dynamics structure. However, we also emphasize that there exists a step-change between one-field and multi-field cases where the considered two-field case may be seen as the simplest example: on the coarse-grained (continuum) level  the one-component systems always approach a stationary state with steady coexisting phases. For multi-field systems with nonreciprocal interactions phases with oscillatory particle densities can be present even on the coarse-grained mean-field level of description (for simulation results see Refs.~\cite{PaPa2023prr,DAGM2023prl}). The resulting possible phase coexistence of uniform and oscillatory phases is expected to widely occur in systems with conservation laws that show large-scale oscillatory instabilities. This makes extensions to multi-field cases particularly interesting.

Restricting our attention (in contrast to \cite{FHKG2023pre}) to purely mass-conserving dynamics, we note that AMB and NRCH model can both be written as
\begin{equation}\label{eq:spurious_var_struct}
	\partial_t \vecg{u} =  \vec{\nabla}\cdot \left( \tens{Q}(\vecg u) \vec{\nabla} \left(\tilde{\tens{M}}(\vecg u)\frac{\delta \tilde{\mathcal{F}}[\vecg u]}{\delta\vecg{u}}\right) \right)\,,
\end{equation}
with $\vecg{u}=(u_1, u_2,\dots,u_N)$, matrices $\tens Q(\vecg u), \tens M(\vecg u) \in \mathbf{R}^{N\times N}$ and a local functional $\tilde{\mathcal{F}}[\vecg u]$. Here, $\tens Q(\vecg u)$ is symmetric and positive definite. The general properties of $\tilde{\tens{M}}$ are discussed further below.

For the specific NRCH model of the main text, $N=2$, $\tilde{\tens{M}}$ is diagonal and constant, while $\tilde{\tensg{Q}}=\tensg 1$. Therefore, $\tilde{\tens{M}}$ and $\vec{\nabla}$ commute and $\tilde{\tens{M}}$ formally takes the role of a possibly indefinite mobility matrix. Note that already this restricted form describes further active, nonreciprocally coupled field theories, when higher order coupling terms in the energy functional $\tilde{\mathcal{F}}$ are considered. E.g., an energetic coupling term $\sim (\vec\nabla u_1)  \cdot (\vec \nabla u_2)$ results in non-reciprocal cross diffusion $\mu_1^{nr}=\alpha \vec\nabla^2 u_2$, $\mu_2^{\text{nr}}=-\alpha \vec\nabla^2 u_1$. Likewise, a higher order coupling energy $\sim u_1^2u_2^2$ results in a nonlinear nonreciprocal interaction $\mu_1^{\text{nr}}=\alpha u_1 u_2^2$ and $\mu_2^{\text{nr}}=-\alpha u_1^2u_2$. However, obviously nonreciprocal interactions that arise from a constant diagonal matrix $\tilde{\tens M}(\vecg u)$ are always of the form $\mu_1^{\text{nr}}=\alpha \delta \mathcal{F}_{12}/\delta u_1$ and $\mu_2^{\text{nr}}=-\alpha \delta \mathcal{F}_{12}/\delta u_2$.

In general, nonconstant $\tilde{\tens{M}}(\vecg u)$ can be incorporated, that do not commute with $\vec \nabla$, which is the case for AMB. The version of AMB presented in \cite{SSCK2018pre} is the one-field model
\begin{equation}\label{eq:aModelB+}
\partial_t u=\vec{\nabla}\cdot(Q(u)\vec{\nabla} g(u))
\qquad\text{with}\qquad g(u)=g_0(u)+\lambda(u)|\vec{\nabla} u|^2-\kappa(u)\vec{\nabla}^2 u \,. 
\end{equation}

In Ref.~\cite{SSCK2018pre}, it is shown\footnote{In particular, in the notation of Eq.~(7) of Ref.~\cite{SSCK2018pre} the dynamical equation is $\dot{\rho}=\vec \nabla \cdot [M[\rho]\vec \nabla \frac{\delta \mathcal{F}}{\delta R}]$ with the scalar field variable $\rho$ and a field transformation $R(\rho)$. The variational chain rule for the functional derivative $\frac{\delta \mathcal{F}}{\delta R}=\frac{\partial \rho}{\partial R}\frac{\delta \mathcal{F}}{\delta \rho}$ and comparison to \eqref{eq:spurious_var_struct} yields that $m(u)$ coincides with $\frac{\partial \rho}{\partial R}$, and $R'=\frac{\partial R}{\partial \rho}$ in Eq.~(4) of \cite{SSCK2018pre} coincides with $\frac{1}{m(u)}$. Therefore, the expressions given here in terms of $m$ slightly differ from those in terms of $R$ in \cite{SSCK2018pre}. In particular, the relation \eqref{eq:def_M_mips2} has a positive sign on the r.h.s. compared to Eq.~(4) of \cite{SSCK2018pre}. } that \eqref{eq:aModelB+} can be written as \eqref{eq:spurious_var_struct} with $N=1$. There, the energy is given as

\begin{equation}\label{eq:def_M_mips}
  \tilde{\mathcal{F}}[u]=\int_\Omega \left[\tilde{f}_h(u)+\frac{\kappa(u)}{2m(u)}|\vec{\nabla} u|^2\right]\mathrm{d}^n x \quad	\text{with}\quad	\tilde{f}_h(u)=\int^u \frac{g_0(\hat{u})}{m(\hat{u})}\mathrm{d}\hat{u}
\end{equation}
where the function $m(u)$ solves the differential equation 
\begin{equation}\label{eq:def_M_mips2}
\kappa m'(u)=\left(2\lambda(u)+\kappa'(u)\right)m(u)
\end{equation}
In other words,  it is given by
\begin{equation}\label{eq:def_M_mips3}
  m(u) =  \exp\left[\int
    \frac{2\lambda(u)+\kappa'(u)}{\kappa(u)}
   \,du\right],
\end{equation}
and $\tilde{\tens{M}}$ corresponds to $m(u)$. 

Coming back to the general form \eqref{eq:spurious_var_struct} we next consider stationary states. They are obtained if one sets $\partial_t u=0$, integrates once, and chooses the integration constant to be zero (i.e., assuming zero net flow across the boundaries). Multiplying by $\tensg Q(\vecg u)^{-1}$ and integrating again one obtains the steady state equation
\begin{equation}\label{eq:twin_general0-u}
	\tilde{\tens M}(\vecg u)\frac{\delta \tilde{\mathcal{F}}[\vecg u]}{ \delta \vecg u}= \vecg \mu = \text{const.},
      \end{equation}
If the matrix $\tilde{\tens M}$ corresponds to the Jacobian of a transformation of variables $\vecg u\to \vecg \chi$, i.e., $\tilde{\tens M}=\partial \vecg u/ \partial \vecg \chi$ then one has
\begin{equation}\label{eq:twin_general0-u}
	\frac{\partial \vecg u}{\partial \vecg \chi} \frac{\delta \tilde{\mathcal{F}}[\vecg u]}{\delta \vecg u}=\frac{\delta \tilde{\mathcal{F}}[\vecg u(\vecg\chi)]}{ \delta \vecg \chi}= \vecg \mu 
\end{equation}
and the stationary states correspond to extrema of the spurious grand potential $\tilde{\Omega}[\vecg \chi]=\tilde{\mathcal{F}}[\vecg\chi]-\vecg\mu \vecg\chi$.

In consequence, one may then for all these models (as done above in section~\ref{sec:MW_NRCH} for the specific NRCH example) define spurious momenta and a spatial Hamiltonian, obtain the generalized spurious pressure $\tilde{p}$ as the third coexistence condition needed for the spurious Maxwell construction. 

Note, that in general, a mobility matrix $\tensg Q(\vecg u)$, even a nonsymmetric or a not positive definite one, does not affect the spurious Maxwell construction for the system \eqref{eq:spurious_var_struct} (as long as $\tensg Q$ is invertible). However, the linear stability thresholds depend on $\tens Q(\vecg u)$, and therefore changing the mobility can render stable coexisting states unstable and vice versa. Finally, note that the additional incorporation of a mass matrix multiplying the time derivatives of the fields on the left hand side of \eqref{eq:spurious_var_struct} in analogy to the  (purely non-mass-conserving case) of the skew-gradient dissipative systems discussed in \cite{KuYa2003pd} does not change the main argument. Although in this way the different approaches seem to converge into the discussed more general form, we expect further mathematical generalizations to exist and to be discovered in the future.

\section{Numerical details, accompanying figures and videos}

The time simulations shown in the main text are computed on 1D or 2D domains with periodic boundary conditions. We separate the linear diagonal part of the r.h.s.\ of Eq.~\eqref{eq:nrch}, i.e., 
\begin{equation}
\begin{split}
	\partial_t \vecg u &= \tensg L_D \vecg u+ \vecg N_L(\vecg u) \\
	\text{where}\quad\tensg L_D &= \begin{pmatrix}
	a\vec \nabla^2+\vec \nabla^4&0\\0& a\vec \nabla^2+\kappa\vec \nabla^4
	\end{pmatrix}\\
	\text{and}\quad \vecg N_L(\vecg u)&=\vec\nabla^2 \begin{pmatrix}
	u_1^3-(\rho+\alpha)u_2\\
	u_2^3-(\rho-\alpha)u_1\\
	\end{pmatrix}.
\end{split}
\end{equation}
We then use a pseudo-spectral method with semi-implicit Euler timestepping where $\tensg L_D$ is treated implicitly and $\vecg N_L(\vecg u)$ is treated explicitly. The timestep is adaptively controlled by a half-step method, where the maximal permitted local error per timestep is usually chosen to be of order $\sim 10^{-5}$ and the resulting timestep usually varies between $10^{-3}$ and $10^{-1}$.

To obtain bifurcation diagrams, we employ the Matlab FEM continuation package \textit{pde2path} \cite{UeWR2014nmma,Ueck2019ccp} and numerically follow steady (temporally periodic) states of Eqs.~\eqref{eq:nrch} with periodic boundary conditions for Figs.~3 and 4 of the main text and with Neumann (homogeneous) boundary condition for Figs.~\ref{fig:bif_ana_passive.pdf}  and \ref{fig:bif_tails}. To follow a specific straight path' through the ($\bar u_1,\bar u_2$)-plane,
we employ the average concentration of species $\bar{u}_1$ as our main continuation parameter and fix the second average concentration, e.g., such as to stay on a tie line or on a straight line through a three-phase region ($\bar{u}_2=m\bar{u}_1+b$ with fixed parameters $m$ and $b$). 
The spatial resolution of Fig.~4 of the main text corresponds to 1024 grid points.  To obtain the branches of time-periodic states in Fig.~3 of the main text, we use a regular grid consisting of  256 spatial grid points and  96 temporal grid points. The linear stability of time-periodic states is calculated using Floquet algorithm~1 of \textit{pde2path} (see section 2.4 of Ref.~\cite{Ueck2019ccp}), where we calculate the 16 largest Floquet multipliers for each state.

\subsection{Additional information for Fig.~1 of the main text}

Table \ref{tab:movies} gives an overview of the movies provided as part of the Supplementary Material that accompany the snapshots from the dynamics highlighted in Fig.~1 of the main text. All simulations are initialized with a phase-separated state where a central circular patch of the high-concentration phase is embedded in a background of the low-concentration phase. Smooth interfaces are used as well as small additional noise of amplitude $\sim 10^{-2}$.  The plateau values correspond to the binodal values as obtained from the spurious Maxwell construction and indicated in Fig.~2 of the main text. For instance, for Fig.~1~(a) of the main text, the initial condition for $u_1$ is given by
\begin{equation}
	u_1(\vec x)=u_{1,\downarrow\downarrow}+(u_{1,\downarrow\uparrow}-u_{1,\downarrow\downarrow})\frac{1}{2}\left[1+\tanh\left(r-|\vec x |\right)\right]+\text{``noise''}
\end{equation}
with $r=25\pi$.
 
\begin{figure}
	\includegraphics[width=\textwidth]{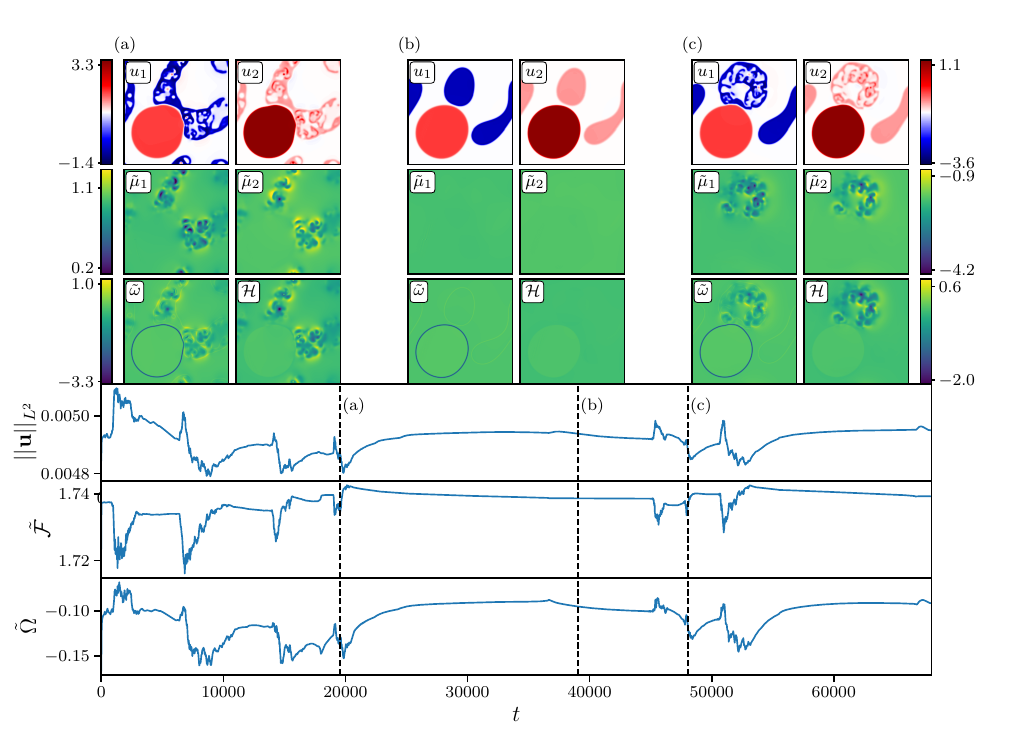}
	\caption{Additional data corresponding to Fig.~1~(c) of the main text (see supplementary movie \textit{1b-three\_phase.mp4}). Panels (a) to (c) additionally show the fields $u_1$ and $u_2$ at three selected times and additionally the spatially resolved spurious thermodynamic quantities $\tilde{\mu}_1$, $\tilde{\mu}_2$, $\tilde{\omega}$ and $\tilde{\mathcal{H}}$ according to Eqs.~\eqref{eq:chem_pot},\eqref{eq:grand_potential} and \eqref{eq:hamiltonian}. The lower row displays the temporal evolution of the $L^2$ norm, the spurious energy $\tilde{\mathcal{F}}$ \eqref{eq:ps_Max_NRCH} and the spurious grand potential $\tilde{\Omega}$.}
	\label{fig:add_ana_fig1b}
\end{figure}

Fig.~\ref{fig:add_ana_fig1b} provides additional information for the three-phase coexistence displayed in Fig.~1~(c) of the main text. Here, the $\uparrow\uparrow$-phase (red) and white $\uparrow\downarrow$-phase are stable, whereas the nonreciprocity is just strong enough to render the third coexisting $\downarrow\downarrow$-phase (blue) weakly oscillatory unstable. Hence, the system is close to a quasi-variational, stationary phase separation, but still includes an active, oscillatory phase. This is reflected in Fig.~\ref{fig:add_ana_fig1b} and the accompanying  movie: For the mayor part of the simulation time, the system shows the usual slow coarsening dynamics known from passive Cahn-Hilliard models, which is reflected in the spurious thermodynamic quantities $\tilde{\mu}_1$, $\tilde{\mu}_2$ and $\tilde{\mathcal{H}}$ being almost spatially constant and the energy $\tilde{\mathcal{F}}$ decreasing slowly. However, occasionally, the active blue phase destabilizes and waves spread in an explosive burst, which is also reflected in the spurious thermodynamic quantities, that strongly fluctuate during the bursts.

\begin{table}
\caption{Summary of movies accompanying Figure 1 of the main text.}
\begin{tabular}{p{0.05\textwidth} | p{0.18\textwidth} p{0.08\textwidth} p{0.12\textwidth} p{0.22\textwidth}| p{0.23\textwidth}}
\toprule
Panel&  corresponding Movie& Param.& domain (grid)& description & remarks\\\midrule
(a)& \textit{1a-cookie\_state.mp4}&$a=-1.5$ \newline$\kappa=1$ \newline$\rho=1$\newline $\alpha=2$&$80\pi\times 80\pi$\newline ($512\times 512$)& cluster of oscillatory phase with irregular wave dynamics &initial transient not shown (video starts at $t=10^4$)\newline corresponds to binodal highlighted in Fig.~2(c)\\[.5em]
(c)&\textit{1b-emergence.mp4}&$a=-1.5$ \newline$\kappa=1$ \newline$\rho=1$\newline $\alpha=2$ &$30\pi\times 30\pi$\newline($256\times 256$)&Emergence of irregular wave dynamics with increasing domain size with radii varying from $4\pi$ to $9\pi$ &corresponds to binodal highlighted in Fig.~2(c)\\[3em]
(b)&\textit{1c-three\_phase.mp4}&$a=-1.5$ \newline$\kappa=1$ \newline$\rho=1$\newline $\alpha=1.3$ &$80\pi\times 80\pi$\newline($512\times 512$)&three-phase coexistence with occasional bursts of irregular wave activity& corresponds to triplepoint highlighted  in Fig.~2(b) \newline additional data see Fig.~\ref{fig:add_ana_fig1b}\\[2em]
(d)&\textit{1d-liquid\_crystal.mp4}&$a=-1.5$ \newline$\kappa=20$ \newline$\rho=1$\newline $\alpha=3$ &$80\pi\times 80\pi$\newline ($1024\times 1024$)&Formation of crystallite (localized state) with hexagonal order&corresponds to binodal in Fig.~2(d)/4(a)\\
\bottomrule
\end{tabular}
\label{tab:movies}
\end{table}

\subsection{Additional information for Fig.~3 of the main text}

\begin{figure*}[h]
\centering
\includegraphics[width=\hsize]{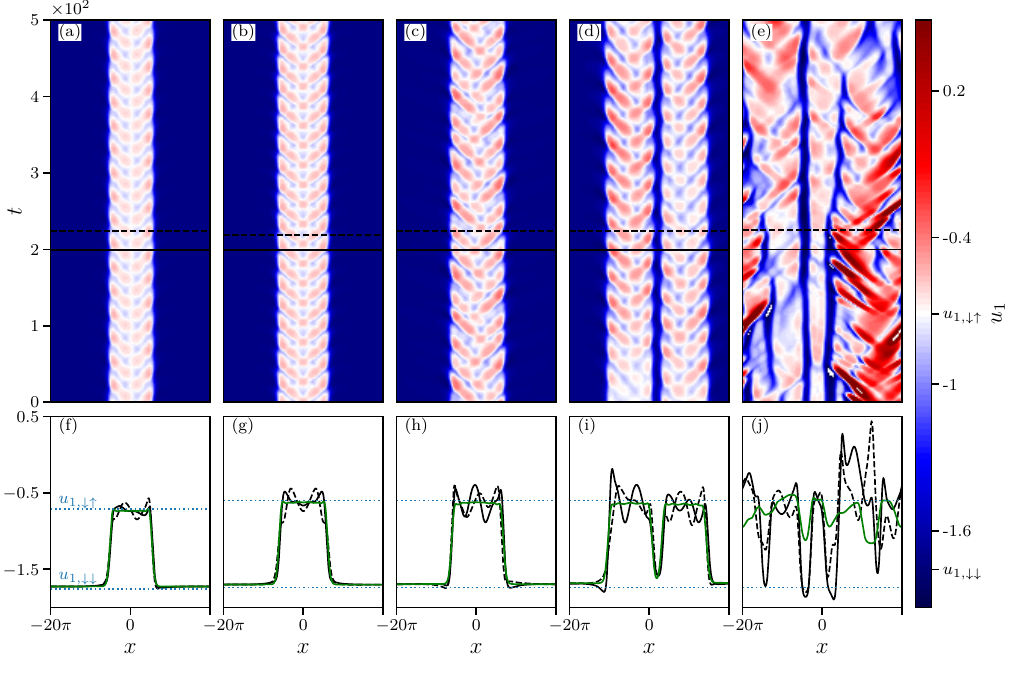}
\caption{Panels (a) to (e) show space-time plots of species $u_1$ at $\bar{u}_1=-1.46$, $\bar{u}_1=-1.42$, $\bar{u}_1=-1.40$, $\bar{u}_1=-1.17$ and $\bar{u}_1=-0.87$, respectively. Snapshots at the respective solid and dashed black horizontal lines are depicted in matching line styles in panels (f) to (j) where the green lines indicate the temporal average and the dotted lines mark the values predicted by the Maxwell construction. The parameters are the same as in Fig.~4 of the main text, i.e. $\bar{u}_2=1.116\bar{u}_1+1.081$. The domain size is $L=40\pi$. The spatial discretization consists of 256 points. }
\label{fig:spacetime}
\end{figure*}

To strengthen the results of the bifurcation analysis presented in Fig.~3  of the main text, we also perform time simulations that are initialized with the unstable stationary phase-separated state with small added noise. 
Fig.~\ref{fig:spacetime} shows spacetime-plots and exemplary profiles of states that emerge after an initial transient ($t>10^5$). The spurious Maxwell construction approximately governs the plateau concentration of the uniform stable phase and the average concentration of the oscillatory phase (Fig.~\ref{fig:spacetime} (f)-(i)).  In a sense, the classical coexistence of uniform stationary phases is replaced by a coexistence of a stationary phase with an oscillatory phase

Close to but beyond the oscillatory instability of the upper plateau at $\bar{u}_1\approx-1.46$ the system settles to the states with simple periodicity (Figs.~\ref{fig:spacetime}(a) and~\ref{fig:spacetime}(b) of the main text), that correspond to the first two time-periodic branches in the bifurcation diagram Fig.~(4)~(b). Increasing $\bar{u}_1$, i.e., increasing the extent of the oscillatory domain leads to intricate  spatio-temporal patterns within this domain (Figs.~\ref{fig:spacetime}(c) and~\ref{fig:spacetime} (d)). Finally, close to the right binodal (Fig.~\ref{fig:spacetime} (e)), where nearly the whole domain is filled by the oscillatory phase, the domain of the uniform stationary phase is reduced to a few ``holes''  and the oscillatory dynamics becomes system filling and rather irregular. 

\section{Data availability}
The data sets and plot functions for all figures as well as examples of
Matlab codes for the employed numerical path continuation are provided on the open source platform zenodo \cite{GLFT2024zenodo}.